# Don't Thrash: How to Cache Your Hash on Flash

Michael A. Bender*§    Martin Farach-Colton†§    Rob Johnson*    Russell Kraner¶
Bradley C. Kuszmaul‡§    Dzejla Medjedovic*    Pablo Montes*    Pradeep Shetty*
Richard P. Spillane*    Erez Zadok*

*{bender,rob,dmededov,pmontes,pshetty,spillane,ezk}@cs.stonybrook.edu
†farach@cs.rutgers.edu    ¶rbkraner@gmail.com    ‡bradley@mit.edu

## ABSTRACT

This paper presents new alternatives to the well-known Bloom filter data structure. The Bloom filter, a compact data structure supporting set insertion and membership queries, has found wide application in databases, storage systems, and networks. Because the Bloom filter performs frequent random reads and writes, it is used almost exclusively in RAM, limiting the size of the sets it can represent.

This paper first describes the quotient filter, which supports the basic operations of the Bloom filter, achieving roughly comparable performance in terms of space and time, but with better data locality. Operations on the quotient filter require only a small number of contiguous accesses. The quotient filter has other advantages over the Bloom filter: it supports deletions, it can be dynamically resized, and two quotient filters can be efficiently merged.

The paper then gives two data structures, the buffered quotient filter and the cascade filter, which exploit the quotient filter advantages and thus serve as SSD-optimized alternatives to the Bloom filter. The cascade filter has better asymptotic I/O performance than the buffered quotient filter, but the buffered quotient filter outperforms the cascade filter on small to medium data sets. Both data structures significantly outperform recently-proposed SSD-optimized Bloom filter variants, such as the elevator Bloom filter, buffered Bloom filter, and forest-structured Bloom filter. In experiments, the cascade filter and buffered quotient filter performed insertions 8.6-11 times faster than the fastest Bloom filter variant and performed lookups 0.94-2.56 times faster.

## 1. INTRODUCTION

Many databases, storage systems, and network protocols maintain Bloom filters [2] in RAM in order to quickly satisfy queries for elements that do not exist in the database, in external storage, or on a remote network host.

The Bloom filter is the classic example of an approximate membership query data structure (AMQ). A Bloom filter supports insert and lookup operations on a set of keys. For a key in the set, lookup returns "present." For a key not in the set, lookup returns "absent" with probability at least $1 - \varepsilon$, where $\varepsilon$ is a tunable false-positive rate. There is a tradeoff between $\varepsilon$ and the space consumption. Other AMQs, such as counting Bloom filters, additionally support deletes [3, 11]. For a comprehensive review of Bloom filters, see Broder and Mitzenmacher [4].

Bloom filters work well when they fit in main memory. However, Bloom filters require about one byte per stored data item. Counting Bloom filters—those supporting insertions and deletions [11]—require 4 times more space [3]. Once Bloom filters get larger than RAM, their performance decays because they use random reads and writes, which do not scale efficiently to external storage, such as flash.

## Results

This paper presents three alternatives to the Bloom filter (BF): the quotient filter (QF), the buffered quotient filter (BQF), and the cascade filter (CF). The QF is designed to run in RAM and the BQF and CF are designed to run on SSD. Unlike the BF, which performs many random writes, these data structures achieve good data locality, and all three support deletions.

The CF is asymptotically more efficient than the BQF at insertions, and thus performs well when the data structure grows much larger than RAM; the BQF is slightly more optimized for queries.

Our evaluation compares the QFs, BQFs, and CFs to BFs and recently proposed BF variants, including buffered Bloom filters (BBF) [5], forest-structured Bloom filters (FBF) [17], and elevator Bloom filters (EBF). For the overview of BF variants, see Section 2. The BBF and FBF were proposed to address the scaling problems of Bloom filters, in particular, when they spill onto SSDs. The EBF is an extension of the BF, which we include as a baseline.

To differentiate the previously existing structures: the EBF is a straightforward application of buffering to BFs. The BBF uses buffering and hash localization to improve

This research was supported in part by DOE Grant DE-FG02-08ER25853, NSF Grants CCF-0540897, CNS-0627645, CCF-0634793, CCF-0937829, CCF-0937833, CCF-0937854, CCF-0937860, and CCF-0937822, and Politécnico Grancolombiano.

SSD performance. The FBF uses buffering, hash localization, as well as in-RAM buffer-management techniques.

Table 1 presents a summary of our experimental results. To put these numbers in perspective, on an Intel X-25M SSD drive, we measured 3,910 random 1-byte writes per second and 3,200 random 1-byte reads per second. Sequential reads run at 261 MB/s, and sequential writes run at 109 MB/s.

We performed three sets of experiments: in RAM, small-scale on SSD, and large-scale on SSD. We performed the different SSD experiments because the effectiveness of buffering decreases as the ratio of in-RAM to on-disk data decreases.

In each case, we compared the rate of insertions, the rate of uniform random lookups, which amounts to lookups for elements not in the AMQ, and the rate of successful lookups, that is, lookups of elements present in the AMQ. We make this distinction in lookups because a BF only needs to check an expected two bits for unsuccessful lookups, but $k$ bits for successful lookups when there are $k$ hash functions. (For our error rates, the BF had 6, 9, and 12 hash functions, respectively.)

*In-RAM Experiments*

For our in-RAM experiments, we compare the QF and the BF. The QF is supposed to be used when it is at most 75% full; as Figure 6 shows, the QF performance deteriorates as it fills. Table 1 reports on results when the structures are 75% full.

For inserts, QFs outperform BFs by factors of 1.3× to 2.5×, depending on the false positive rates. For uniform random lookups, BFs are 1.4×-1.6× faster. For successful lookups, there is no clear winner.

*Small On-SSD Experiments*

We compared our two SSD data structures to the three Bloom filter variants. In these experiments, the AMQs were grown so that they are approximately four times the size of RAM. See Section 5.2 for details.

We find that both BQF and CF insert at least 4 times faster than other data structures and that BQF is at least twice as fast for lookups as all the other AMQs we measured. In fact, on successful lookups, it runs roughly 11 times better than EBF and BBF.

The BQF is the clear winner for this set of experiments.

*Large On-SSD Experiments*

We ran all AMQs for 35,000 seconds. This was enough time for CF and BQF to insert the full data set. However, BBF, FBF, and EBF were at least 10 times slower for insertions and none of them managed to get through even 10% of the insertion load. We therefore conclude that these data structures are not suitable for such workloads.

We note that this workload was large enough for asymptotics to kick in: the CF was 26% faster than the BQF. BQF still dominates for queries, outperforming CF by at least 60%. Therefore the choice of CF versus BQF depends on the ratio of insertions to queries in a particular workload.

*Other Considerations*

For typical configurations, e.g. a 1% false positive rate, a QF uses about 20% more space than a BF. However, QFs (and BQFs and CFs) support deletion, whereas BFs incur a 4× space blow-up to support deletion, and even then they may fail. QFs support in-order iteration over the hash values inserted into the filter. Consequently, QFs can be dynamically resized, and two QFs can be merged into a single larger filter using an algorithm similar to the merge operation in merge sort. QF inserts and lookups require a single random write or read. BF inserts require multiple writes, and lookups require two reads on average.

## Applications

Write-optimized AMQs, such as the CF and BQF, can provide a performance improvement in databases in which inserts and queries are *decoupled*, i.e. insertion operations do not depend on the results of query operations. Webtable [6], a database that associates domain names of websites with website attributes, exemplifies such a workload. An automated web crawler adds new entries into the database while users independently perform queries. The Webtable workload is decoupled because it permits duplicate entries, meaning that searches for duplicates need not be performed before each insertion.

The system optimizes for a high insertion rate by splitting the database tables into smaller subtables, and searches are replicated across all the subtables. To make searches fast, the system maintains an in-memory Bloom filter for each subtable. The Bloom filter enables the database to avoid I/O to subtables that do not contain the queried element.

The CF and BQF could enable databases, such as Webtable, to scale to larger sizes without a concomitant increase in RAM. SSD-optimized AMQs, such as the CF and BQF, can keep up with the high insertion throughput of write-optimized databases.

Similar workloads to Webtable, which also require fast insertions and independent searches, are growing in importance [6, 12, 14]. Bloom filters are also used for deduplication [22], distributed information retrieval [20], network computing [4], stream computing [21], bioinformatics [7, 18], database querying [19], and probabilistic verification [13].

The remainder of this paper is organized as follows. Section 2 describes the Bloom filter and its external-memory variants. Section 3 presents the quotient filter and gives a theoretical analysis. Section 4 presents the buffered quotient filter and cascade filter. Section 5 presents our experiments.

## 2. BLOOM FILTER AND SSD VARIANTS

This section reviews the traditional Bloom filter and its SSD variants.

A Bloom filter $B$ is a lossy, space-efficient representation of a set. It supports two operations: INSERT$(B, x)$ and MAY-CONTAIN$(B, x)$.

A BF $B$ consists of a bit array $B[0..m-1]$ and $k$ hash functions $h_i : \mathcal{U} \to \{0, \ldots, m-1\}$, where $1 \leq i \leq k$ and $\mathcal{U}$ is the universe of objects that may be inserted into the filter. To insert an item $x$, the filter sets

$$B[h_i(x)] \leftarrow 1 \qquad \text{for } i = 1, \ldots, k.$$

To test whether an element $x$ may have ever been inserted, the filter checks all the bits that would have been set:

$$\text{MAY-CONTAIN}(B, x) = \bigwedge_{i=1}^{k} B[h_i(x)].$$

(a) In-RAM experimental results (operations per second).

| AMQ | BF | QF | BF | QF | BF | QF |
|---|---|---|---|---|---|---|
| **False Positive Rate** | **0.01** | **0.01** | **0.002** | **0.002** | **0.0002** | **0.0002** |
| **Uniform Random Inserts** | 1.72 mil | 2.44 mil | 1.29 mil | 2.43 mil | 991,000 | 2.45 mil |
| **Uniform Random Lookups** | 3.1 mil | 2.1 mil | 3.35 mil | 1.98 mil | 3.37 mil | 2.13 mil |
| **Successful Lookups** | 1.93 mil | 1.61 mil | 1.65 mil | 1.7 mil | 1.44 mil | 1.71 mil |

(b) On-disk experimental results (operations per second).

| AMQ | | CF | BQF | EBF | BBF | FBF |
|---|---|---|---|---|---|---|
| **Small experiment** | Uniform Random Inserts | 1.075 mil | 1.32 mil | 205,000 | 249,000 | 43,100 |
| | Uniform Random Lookups | 2,200 | 4,480 | 2,180 | 2,340 | 1,510 |
| | Successful Lookups | 2,950 | 4,690 | 372 | 441 | 1,830 |
| **Large experiment** | Uniform Random Inserts | 728,000 | 576,000 | | | |
| | Uniform Random Lookups | 1,940 | 3,600 | | | |
| | Successful Lookups | 2,380 | 3,780 | | | |

Table 1: Summary of evaluation results.

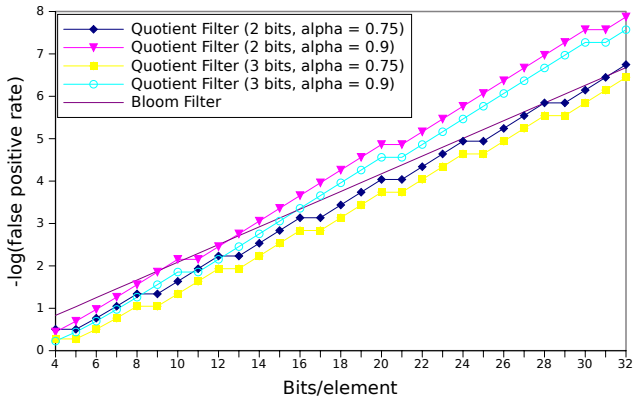

Figure 1: False positive rates for BF and QF. For typical parameters (e.g., 1% false positive rate), QF require about 20% more space than a BF. For extremely low false positive rates, QF use less space than a BF.

The false-positive rate of a BF after inserting $n$ items is approximately

$$(1 - e^{-nk/m})^k.$$

This rate is optimized by choosing

$$k = \frac{m}{n} \ln 2,$$

which means that roughly half of the bits in $B$ are set to 1.

As a concrete example, an optimally filled BF with $m = 8n$ (i.e., 1 byte per element) would use six hash functions and can achieve a false positive rate of 1.56%. Figure 2 shows the BF false positive rate, assuming the optimal number of hash functions, as a function of the number of bits per element.

BFs has several limitations. A BF does not expand to accommodate new elements, so sufficient space for all the elements must be allocated in advance. A BF does not support deletions. A BF does not naturally scale to external storage because of the poor data locality, and consequently they are usually stored in RAM. To illustrate, a BF stored on a rotating disk, with $k = 10$ hash functions, could insert fewer than 20 elements per second.

Researchers have devised several approaches to improve BF scalability:

- *Replacing magnetic disks with SSDs.* SSDs offer random read and write rates superior to those of magnetic disks. With an off-the-shelf SSD, the traditional BF with $k = 10$ hash functions can achieve roughly 500 inserts per second. High-end devices, such as FusionIO [8], can offer further speedups.

- *Buffering.* Reserve a buffer space in RAM, and cache these updates in the buffer. Flush the buffer as it becomes full. With buffering, multiple bit writes destined for the same SSD block require only one I/O. The elevator Bloom filter implements this strategy. In general, buffering performs well when the ratio between the Bloom filter size and the RAM buffer size is small. As described in [5], queries can also be delayed and buffered in a multithreaded environment, but the present paper measures the performance when queries must be answered immediately.

- *Hash localization.* Improve data locality by directing all hashes of one insertion into a single SSD block. When combined with buffering, this can substantially improve the locality of writes. Queries see a less dramatic improvement in locality. BF variants, such as the buffered Bloom filter [5] and the closely related BloomFlash [10], use this strategy.

- *Multi-layered design.* Maintain multiple on-disk BFs, exponentially increasing in size. Insert only into the largest and most recent BF. This approach effectively reduces the ratio between the RAM size and the active BF by a factor of 2, but increases the search cost, since a search must query all Bloom filters. The forest-structured Bloom filter [17] uses this strategy.

- *Buffer design and flushing policy.* Different buffer management schemes may lead to different performance characteristics. In the BBF, the buffer is equally divided into a number of sub-buffers, each serving updates for a particular SSD block. When a sub-buffer becomes full, its updates are applied with one I/O. BloomFlash flushes the group of $c$ contiguous sub-buffers that has the most updates, and optimizes for $c$.

## 3. QUOTIENT FILTER

In this section we describe the quotient filter, a space-efficient and cache-friendly data structure that delivers all the functionality of the Bloom filter. We explain advantages of the QF over the BF that make the QF particularly suitable to serve as the foundation for our SSD-resident data structures. Finally, we give implementation details, describe potential variations, and analyze asymptotic performance.

The QF represents a multi-set of elements $S \subseteq \mathcal{U}$ by storing a $p$-bit fingerprint for each of its elements. Specifically, the QF stores the multi-set $F = h(S) = \{h(x) \mid x \in S\}$, where $h : \mathcal{U} \to \{0, \ldots, 2^p - 1\}$ is a hash function. To insert an element $x$ into $S$, we insert $h(x)$ into $F$. To test whether an element $x \in S$, we check whether $h(x) \in F$. To remove an element $x$ from $S$, we remove (one copy of) $h(x)$ from $F$.

Conceptually, we can think of $F$ as being stored in an open hash table $T$ with $m = 2^q$ buckets using a technique called *quotienting*, suggested by Knuth [15, Section 6.4, exercise 13]; see the open hash table (i.e., hash table with chaining) at the top of Figure 2. In this technique a fingerprint $f$ is partitioned into its $r$ least significant bits, $f_r = f \bmod 2^r$ (the remainder), and its $q = p - r$ most significant bits, $f_q = \lfloor f/2^r \rfloor$ (the quotient). To insert a fingerprint $f$ into $F$, we store $f_r$ in bucket $T[f_q]$. Given a remainder $f_r$ in bucket $f_q$, the full fingerprint can be uniquely reconstructed as $f = f_q 2^r + f_r$.

To reduce the memory required to store the fingerprints and achieve better spatial locality, the hash table is compactly stored in an array $A[0 \ldots m-1]$ of $(r+3)$-bit items, similar to that described by Cleary [9]; see Figure 2, bottom. Each slot in $A$ stores an $r$-bit remainder along with three meta-data bits, which enable perfect reconstruction of the open hash table.

If two fingerprints $f$ and $f'$ have the same quotient ($f_q = f'_q$) we say there is a *soft collision*. In this case we use linear probing as a collision-resolution strategy. All remainders of fingerprints with the same quotient are stored contiguously in what we call a *run*. If necessary, a remainder is shifted forward from its original location and stored in a subsequent slot, wrapping around at the end of the array. We maintain the invariant that if $f_q < f'_q$, $f_r$ is stored before $f'_r$ in $A$, modulo this wrapping.

The three meta-data bits in each slot of $A$ work as follows. For each slot $i$, we maintain an *is-occupied* bit to quickly check whether there exists a fingerprint $f \in F$ such that $f_q = i$. For a remainder $f_r$ stored in slot $i$, we record whether $f_r$ belongs to bucket $i$ (i.e., $f_q = i$) with an *is-shifted* bit. Finally, for a remainder $f_r$ stored in slot $i$, we keep track of whether $f_r$ belongs to the same run as the remainder stored in slot $i - 1$ with an *is-continuation* bit. Intuitively, the *is-shifted* bit, when it is set to 0, tells the decoder the exact location of a remainder in the open hash table representation, the *is-continuation* bit enables the decoder to group items that belong to the same bucket (runs), and the *is-occupied* bit lets the decoder identify the correct bucket for a run.

We define a *cluster* as a sequence of one or more consecutive runs (with no empty slots in between). A cluster is always immediately preceded by an empty slot and its first item is always un-shifted. The decoder only needs to decode starting from the beginning of a cluster. However, rather than decoding, we can perform all operations in place. Figure 3 shows the algorithm for testing whether a fingerprint $f$ might have been inserted into a QF $A$.

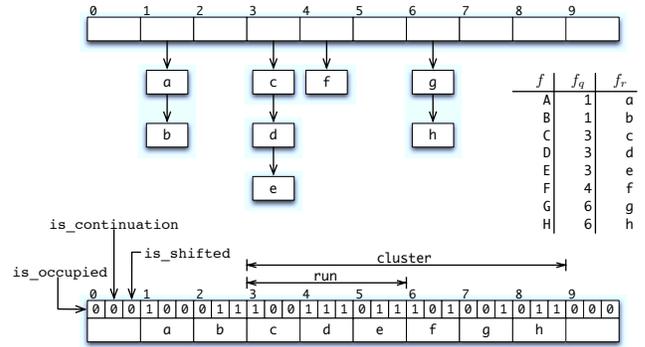

**Figure 2: An example quotient filter with 10 slots along with its equivalent open hash table representation. The remainder, $f_r$, of a fingerprint $f$ is stored in the bucket specified by its quotient, $f_q$. The quotient filter stores the contents of each bucket in contiguous slots, shifting elements as necessary and using three meta-data bits to enable decoding.**

To insert/delete a fingerprint $f$, we operate in a similar manner: first we mark/unmark $A[f_q]$ as occupied. Next, we search for $f_r$ using the same algorithm as MAY-CONTAIN to find the slot where it should go. Finally, we insert/remove $f_r$ and shift subsequent items as necessary, while updating the other two meta-data bits. We stop shifting items as soon as we reach an empty slot.

Since the QF is just a compact representation of $F$, its false positive rate is a function of the hash function, $h$, and the number of items, $n$, inserted into the filter. In particular, a false positive happens when an element $x' \notin S$ has the same fingerprint as an element $x \in S$ ($h(x) = h(x')$). We refer to this event as a *hard collision*. Assuming $h$ generates outputs uniformly and independently distributed in $\{0, \ldots, 2^p - 1\}$, the probability of a hard collision is given by

$$1 - \left(1 - \frac{1}{2^p}\right)^n \approx 1 - e^{-n/2^p} \leq \frac{n}{2^p} \leq \frac{2^q}{2^p} = 2^{-r}.$$

Figure 2 shows the false positive rate (on a log scale) for QF as a function of the bits per element. In the figure, $\alpha$ is the load factor of the QF (i.e., the fraction $n/m$ of occupied slots). Figure 2 also shows the false-positive rate for the BF and for a QF variant, described later, that uses only two meta-data bits per slot.

The time required to perform a lookup, insert, or delete in a QF is dominated by the time to scan backwards and forwards. One such operation need only scan through one cluster. Therefore, we can bound the cost by bounding the size of clusters. The following theorem can be proved by a straightforward application of Chernoff Bounds.

**Fact.** *Let $\alpha \in [0, 1)$. Suppose there are $\alpha m$ items in a quotient filter with $m$ slots. Let*

$$k = (1 + \varepsilon) \frac{\ln m}{\alpha - \ln \alpha - 1}.$$

MAY-CONTAIN($A, f$)
    $f_q \leftarrow \lfloor f/2^r \rfloor$     ▷ quotient
    $f_r \leftarrow f \bmod 2^r$     ▷ remainder
    **if** $\neg$ *is-occupied*($A[f_q]$)
        **then return** FALSE
    ▷ walk back to find the beginning of the cluster
    $b \leftarrow f_q$
    **while** *is-shifted*($A[b]$)
        **do** DECR($b$)
    ▷ walk forward to find the actual start of the run
    $s \leftarrow b$
    **while** $b \neq f_q$
        **do** ▷ invariant: $s$ points to first slot of bucket $b$
            ▷ skip all elements in the current run
            **repeat** INCR($s$)
                **until** $\neg$ *is-continuation*($A[s]$)
            ▷ find the next occupied bucket
            **repeat** INCR($b$)
                **until** *is-occupied*($A[b]$)
    ▷ $s$ now points to the first remainder in bucket $f_q$
    ▷ search for $f_r$ within the run
    **repeat**  **if** $A[s] = f_r$
               **then return** TRUE
            INCR($s$)
    **until** $\neg$ *is-continuation*($A[s]$)
    **return** FALSE

**Figure 3: Algorithm for checking whether a fingerprint $f$ is present in the QF $A$.**

Then

$$\Pr[\text{there exists a cluster of length} \geq k] < m^{-\varepsilon} \xrightarrow{m \to \infty} 0.$$

For example, with $q = 40$ ($m = 2^{40}$) and $\alpha = 3/4$, the largest cluster in the QF has approximately 736 slots. On average, clusters are $O(1)$ in size. The expected length of a cluster is less than $1/(1 - \alpha e^{1-\alpha})$. For example, with $\alpha = 3/4$, the average cluster length is 27. Figure 4 shows the distribution of cluster sizes for three choices of $\alpha$. With $\alpha = 1/2$, 99% of the clusters have less than 24 elements.

We have shown that QF offers space and false-positive performance that is comparable to BF, but QF has several significant advantages.

**Cache friendliness.** QF lookups, inserts, and deletes require decoding and possibly modifying a single cluster. Since clusters are small, these slots usually fit in one or two cache lines. On SSD, they usually fit in one disk page, which can be accessed with a single serial read or write. BF inserts, on the other hand, require writing to $k$ random locations, where $k$ is the number of hash functions used by the filter. Similarly, BF lookups require about two random reads on average for absent elements and $k$ for present elements.

**In-order hash traversal.** As mentioned before, it is possible to reconstruct the exact multi-set of fingerprints inserted into a QF. Furthermore, the QF supports in-order traversal of these fingerprints using a cache-friendly linear scan of the slots in the QF. These two features enable two other useful operations that are not possible with BF: resizing and merging.

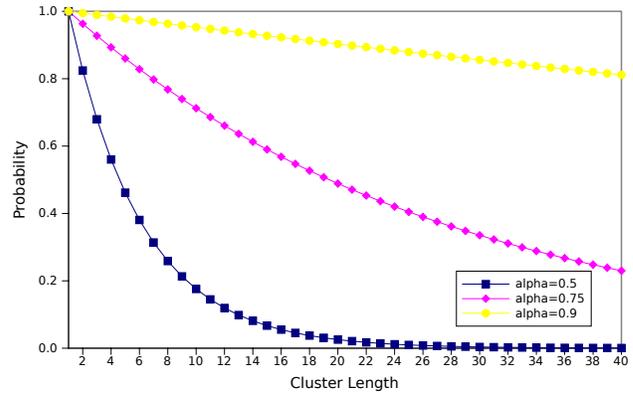

**Figure 4: Distribution of cluster sizes for 3 choices of $\alpha$.**

**Resizing.** Like most hash tables, the QF can be dynamically resized—both expanded and shrunk—as items are added or deleted. Unlike hash tables, however, this can be accomplished without the need of rehashing by simply borrowing/stealing one bit from the remainder into the quotient. This can be implemented by iterating over the array while copying each fingerprint into a newly allocated array.

**Merging.** Similarly, two or more QF can be merged into a single, larger filter using an algorithm similar to that used in merge sort. The merge uses a sequential scan of the two input filters and sequentially writes to the output filter and hence is cache friendly.

**Deletes.** The QF supports correct deletes while standard Bloom Filters do not. In contrast, Counting Bloom filters [3, 11] support probabilistically correct deletions by replacing each bit in a BF with a 4-bit counter, but this incurs a large space overhead and there is still a probability of error.

## Quotient Filter Variants

We now give space-saving variations on the QF. The QF decoder maintains two pointers: $b$, a pointer to the current bucket and $s$, a pointer to the current slot. The decoder needs to initialize $b$ and $s$ to correct values in order to begin decoding. That is the purpose of the *is-shifted* bit: if $\neg$ *is-shifted*($A[i]$), then the decoder can initialize $b = s = i$. There are other ways to initialize $b$ and $s$:

- Synchronizers. The QF could store a secondary array, $S[0 \ldots (2^q/\ell) - 1]$, of $c$-bit items. Entry $S[i]$ would hold the offset between bucket $i\ell$ and the slot holding its first element. The decoder can initialize $b = i\ell$ and $s = i\ell + S[i] \bmod 2^q$ for any $i$. For example, to lookup an element in bucket $f_q$, the decoder would choose $i = \lfloor f_q/\ell \rfloor$. As a special case, when $S[i] = 2^c - 1$, the offset between bucket $i\ell$ and the slot holding its first element is greater than or equal to $2^c - 1$. The decoder cannot use such entries to begin decoding – it must walk backwards to find the nearest index $i$ such that $S[i] < 2^c - 1$. Since clusters are small, so are the offsets, so we can choose small $c$ (e.g., 5 or 8). The frequency, $\ell$, of synchronizers can trade space for decoding speed. The current system, with *is-shifted* bits, is essentially a special case of this scheme with $c = \ell = 1$. By choosing a large $\ell$, the per-slot overhead of the QF can be arbitrarily close to two bits.

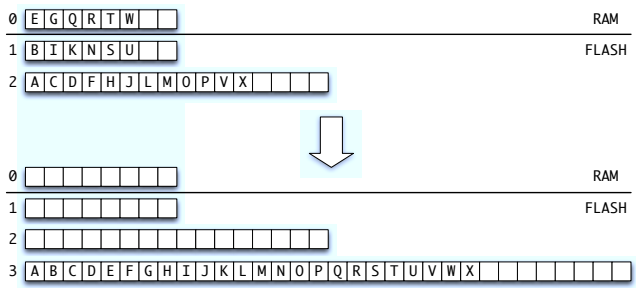

**Figure 5: Merging QFs.** Three QFs of different sizes are shown above, and they are merged into a single large quotient filter below. The top of the figure shows a CF before a merge, with one QF stored in RAM, and two QFs stored in flash. The three QFs above have all reached their maximum load factors (which is 3/4 in this example). The bottom of the figure shows the same CF after the merge. Now the QF at level 3 is at its maximum load factor, and the QFs at levels 0, 1, and 2 are empty.

- Reserved remainders. We can also reserve a special remainder value, e.g. 0, to indicate that a slot is empty, and decoding can begin at an empty slot $i$ with $b = s = i$. This would require only 2 meta-data bits, but reduces the hash space slightly.

- Sorting tricks. Finally, it is possible to indicate empty slots by ordering elements within each bucket and placing "illegal" unordered sequences of elements in empty regions of the QF. In this way, we can achieve exactly two bits of overhead. Decoding in this version is complex and slower.

## 4. QUOTIENT FILTERS ON FLASH

In this section we give two AMQs designed for SSD, the buffered quotient filter and the cascade filter. Both structures use the QF as a building block. The false positive rates of these structures are exactly the same as that of a single QF storing all of the elements.

### Buffered Quotient Filter

The BQF uses one QF as the buffer and another QF on the SSD. When the in-RAM QF becomes full, we sequentially iterate over it and flush elements to disk. The QF serves well as a buffer because of its space efficiency and because it allows the flush to iterate sequentially through its fingerprints and write to SSD. Since elements are stored in sequential order, the writes to SSD will also be sequential. Since each flush may write to every page of the on-disk structure, the amortized cost of inserting an item into a BQF of $n$ items with a cache of size $M$ and a block size of $B$ bytes is $O(\frac{n}{MB})$. The BQF is optimized for lookup performance. Most lookups perform one I/O. As with the buffering approaches from Section 2, performance degrades as the filter-to-RAM size increases.

### Cascade Filter

The CF is optimized for insertion throughput but offers a tradeoff between lookup and insertion speed.

The overall structure of the CF is loosely based on a data structure called the COLA [1]; see Figure 5. The CF maintains an in-memory QF, $Q_0$. In addition, for RAM of size $M$, the CF maintains $\ell = \log(n/M) + O(1)$ in-flash QFs, $Q_1, \ldots, Q_\ell$, of exponentially increasing size. New items are initially inserted into $Q_0$. When $Q_0$ reaches its maximum load factor, the CF finds the smallest $i$ such that the elements in $Q_0, \ldots, Q_i$ can be merged into level $i$. It then creates a new, empty quotient filter $Q'_i$, merges all the elements in $Q_0, \ldots, Q_i$ into $Q'_i$, replaces $Q_i$ by $Q'_i$, and replaces $Q_0, \ldots, Q_{i-1}$ with empty QFs. To perform a CF lookup, we perform a lookup in each nonempty level, which requires fetching one page from each.

It is possible to implement this scheme with different branching factors, $b$. That is, $Q_{i+1}$ can be $b$ times as large as $Q_i$. As $b$ increases, the lookup performance increases because there are fewer levels, but the insertion performance decreases because each level may be rewritten multiple times.

The theoretical analysis of CF performance follows from the COLA: a search requires one block read per level, for a total of $O(\log(n/M))$ block reads, and an insert requires only $O((\log(n/M))/B)$ amortized block writes/erases, where $B$ is the natural block size of the flash. Typically, $B \gg \log(n/M)$, meaning the cost of an insertion or deletion is much less than one block write per element. Like a COLA, a CF can be deamortized to provide better worst-case bounds [1]. This deamortization removes delays caused by merging large QFs.

## 5. EVALUATION

This section answers the following questions:

1. How does the quotient filter compare to the Bloom filter with respect to in-RAM performance?

2. How do the cascade filter and buffered quotient filter compare to various Bloom filter alternatives on Flash?

3. How does the on-disk performance of the cascade filter and buffered quotient filter change as the database scales out of RAM?

4. How do the different data structures compare on lookup performance? We investigate the performance of both successful lookups and uniform random lookups (which are almost all unsuccessful).

5. What is the insert/lookup tradeoff for the cascade filter with varying fan-outs?

This section comprises three parts.

In the first part, we compare the QF and the BF in RAM. We compare the two data structures for three different false positive rates: $1/64 \approx 1\%$, $1/512 \approx 0.2\%$, and $1/4096 \approx 0.02\%$.

In the second part, we measure the on-disk performance of the CF, the BQF, the EBF, the BBF and the FBF. Here, we perform experiments with the RAM-to-database size ratios of 1 : 4 and 1 : 24, which we call small and large experiments, respectively.

In the third part, we measure the performance tradeoffs between the insertion and the lookup performance when varying the fanout of the CF. We report results for fanouts of 2, 4, and 16.

In all experiments, we measure three performance aspects:

**Uniform random inserts:** Keys are selected uniformly from a large universe.

**Uniform random lookups:** Keys are selected as before. When performed on an optimally filled AMQ data structure, such queries will report true with probability equal to that of our false positive rate.

**Successful lookups:** Keys are chosen uniformly at random from one of the keys actually present.

We use an interleaved workload. Every 5% of completed insertions, we spend 60 seconds performing uniform random lookups, followed by 60 seconds performing successful lookups. This way, we can measure the lookup performance at different points of data structure occupancy.

*Experimental Setup.* We created C++ implementations of all the data structures evaluated in these experiments. Our BF, EBF, BBF, and FBF implementations always uses the optimal number of hash functions. The BBF "page size" parameter controls the amount of space that will be written when buffered data is flushed to SSD. We configured our BBF to use 256KB pages, which is the erasure block size on our SSDs, as recommended by the BBF authors. The analogous FBF parameter is called the "block size", and we configured our FBF implementation to use 256KB blocks. The FBF "page size" governs the size of reads performed during lookups; our FBF implementation used 4KB pages.

Our benchmarking infrastructure generated a 512-bit hash for each item inserted or queried in the data structure. Each data structure could partition the bits in this hash as it needed. For example, a BF configured to use 12 hash functions, each with a 24-bit output, would use 288 bits of the 512-bit hash and discard the rest. We chose 512-bit hashes because many real-world AMQ applications, such as de-duplication services, use cryptographic hashes, such as SHA-512.

We ran our experiments on two identically configured machines, running Ubuntu 10.04.2 LTS. Each machine includes a single-socket Intel Xeon X5650 (6 cores with 1 hyperthread on each core, 2.66GHz, 12MB L2 cache). The machines have 64GB of RAM; to test the out-of-RAM performance, we booted them with 3GB each.

Each machine has a 146.2GB 15KRPM SAS disk used as the system disk and a 160GB SATA II 2.5in Intel X25-M Solid State Drive (SSD) used to store the out-of-RAM part of the data. We use only a 95GB partition of the SSD to minimize the SSD FTL firmware interference. We formatted the 95GB partition as an ext4 filesystem and out-of-RAM data was stored in a 80GB file in that filesystem. We used `dd` to zero the file between each experiment. With this configuration, we could perform 3,910 random 1 byte writes per second, 3,200 1 byte random reads per second, sequential reads at 261 MB/s, and sequential writes at 109 MB/s.

To avoid swapping, we set the Linux swappiness to zero and we monitored `vmstat` output to ensure that no swapping occured.

### 5.1 In-RAM Performance: Quotient Filter vs. Bloom Filter

This section presents the experimental comparison of QF to the BF, with varying false positive rates.

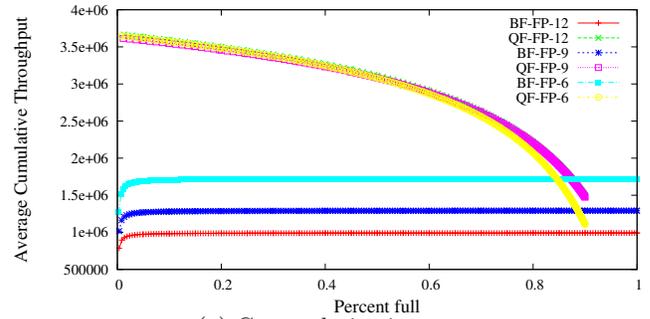

(a) Cummulative inserts

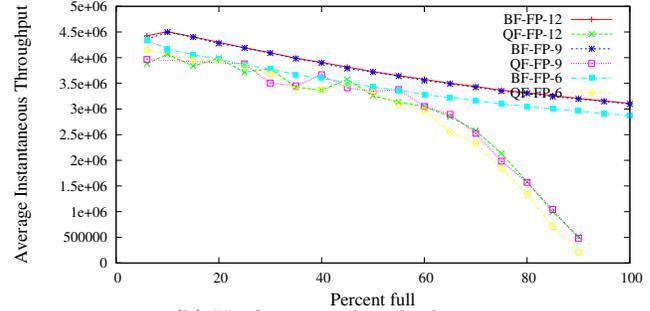

(b) Uniform random lookups

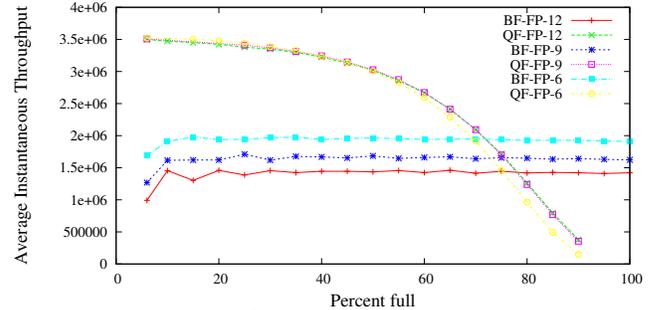

(c) Successful lookups

**Figure 6:** In-RAM Bloom Filter vs. Quotient Filter Performance.

Both data structures were given 2GB of space in RAM and we tested their performance on three false positive rates: 1/64, 1/512, and 1/4096.

In both experiments, we construct the data structures that can fit the maximum number of elements without violating the false positive rate nor the space requirements. We fill the BF to the maximum occupancy. Because the insertion throughput of the QF significantly deteriorates towards maximum occupancy, we let the QF experiment run up to 90% full.

*Results.* Figure 6 shows the insertion, random lookup, and successful lookup throughputs of the BF and quotient filter.

The quotient filter substantially outperforms the BF on insertions until the quotient filter is 80% full. The BF insertion throughput is independent of its ocupancy, but degrades as the false positive rate goes down, since it has to set more bits for each inserted item. The quotient filter insertion throughput is unaffected by the false positive rate, but it gets slower as it becomes full, since clusters become larger.

|          | Capacity     |              |
| FP rate  | BF           | QF (90%)     |
| -------- | ------------ | ------------ |
| 1/64     | 1.98 billion | 1.71 billion |
| 1/512    | 1.32 billion | 1.29 billion |
| 1/4096   | 991 million  | 1.03 billion |

**Table 2: Capacity of the quotient filter and BF data structures used in our in-RAM evaluation. In all cases, the data structures used 2GB of RAM.**

The quotient filter matches the BF random lookup performance until about 65% occupancy. The quotient filter performance degrades as its occupancy increases because clusters become longer. The BF performance degrades because the density of 1 bits increases, so the lookup algorithm must, on average, check more bits before it can conclude that an element is not present.

The quotient filter significantly outperforms the BF on successful lookups up to about 75% capacity. The BF performance is independent of occupancy since, in all successful lookups, it must check the same number of bit positions. The quotient filter performance degrades as clusters get larger.

Table 2 shows the capacity of the BFs and quotient filters in our experiments. As predicted in Figure 2, the capacities are almost identical, with the quotient filter more efficient for lower false positive rates.

Overall, the quotient filter outperforms the BF until its occupancy reaches about 70%. The quotient filter requires slightly more space for high false positive rates, and less space for lower false positive rates.

## 5.2 On-disk Benchmarks

We evaluate the insert and lookup performance of CFs, BQFs, EBFs, BBFs and FBFs when they are bigger than RAM. To see how performance of various data structures scales as the RAM-to-filter ratio shrinks, we run two experiments, with RAM-to-filter ratios of 1 : 4 and 1 : 24. The false positive rate in both experiments is fixed to $f = 1/4096 \approx 0.024\%$, which sets the number of hash functions for the EBF, the BBF and the FBF to $k = 12$, $k = 13$ and $k = 14$, respectively.

We refer to the first experiment, which uses a RAM-to-filter ratio of 1 : 4, as the *small* experiment. The RAM buffer size is set to 2GB and the size of data structures on disk is roughly 8GB. The remaining 1GB of RAM is left for the operating system (to use partly as page cache). We inserted 3.97 billion elements into each data structure.

The second experiment, using a RAM-to-filter ratio of 1 : 24 can be thought of as a "large" experiment. In this case all data structures employ 2GB of RAM buffer, and a 48GB on-disk data structure. As in the previous experiment, 1GB is set aside for the page cache. In this configuration, the CF and BQF can hold 23 billion elements, and they can insert them in under 35,000 seconds. All the other data structures were too slow to complete the experiment – we present only partial results obtained after inserting elements for 35,000 seconds.

*Results.* Figures 7 and 8 show the insertion, random lookup, and successful lookup performance obtained in the small and large experiments. The small CF and BQF ex-periments completed in about 1 hour. The small EBF and BBF experiments took about 10 hours, and the small FBF experiment took about 25 hours to complete. Consequently, Figure 7(a) only shows the throughput of each data structure through the first hour of the small experiment. See Table 1 for the overall throughputs.

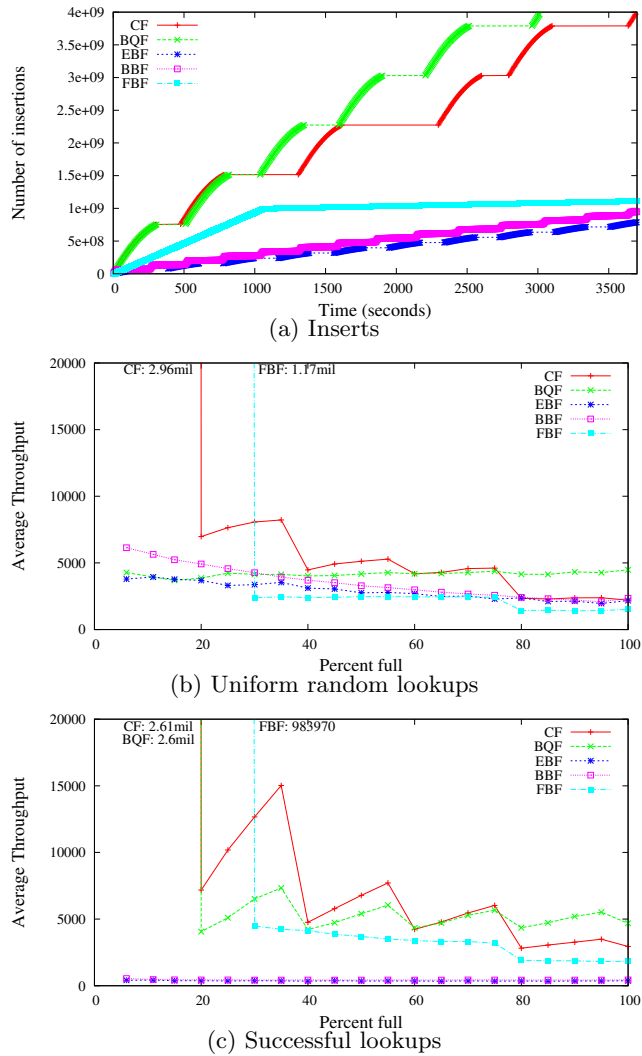

(a) Inserts

(b) Uniform random lookups

(c) Successful lookups

**Figure 7: Small disk experiment. In 7(a), the staircase pattern of the CF is due to the merges of the small QFs into a larger quotient filter. The stalls in the BQF performance are due to flushing of the in-RAM quotient filter to the on-disk quotient filter. In 7(b) and 7(c), the lookup performance of the cascade filter depends on the number of full QFs. The BBF and the EBF perform more poorly on the successful lookups, as they need to check 12 bits, performing roughly 12 random reads.**

In the large experiment, the other three data structures all completed less than 10 percent of the experiment. Figure 8(a) shows their cumulative throughput for the first 35,000 seconds, but Figures 8(b) and 8(c) do not plot their lookup performance, since the data structures were too slow to obtain this data.

There are two main trends to notice in the insertion throughput graphs: (1) the CF and BQF are orders of magnitude faster than the EBF, BBF, and FBF, and (2) the CF scales better than the BQF. In the small experiment, the BQF outperforms the best BF variant by a factor of 5.2, and slightly outperforms the CF. In the large experiment, the CF performs 11 times more insertions than any of the BF variants, and the BQF performs 9 times more insertions than the BF variants.

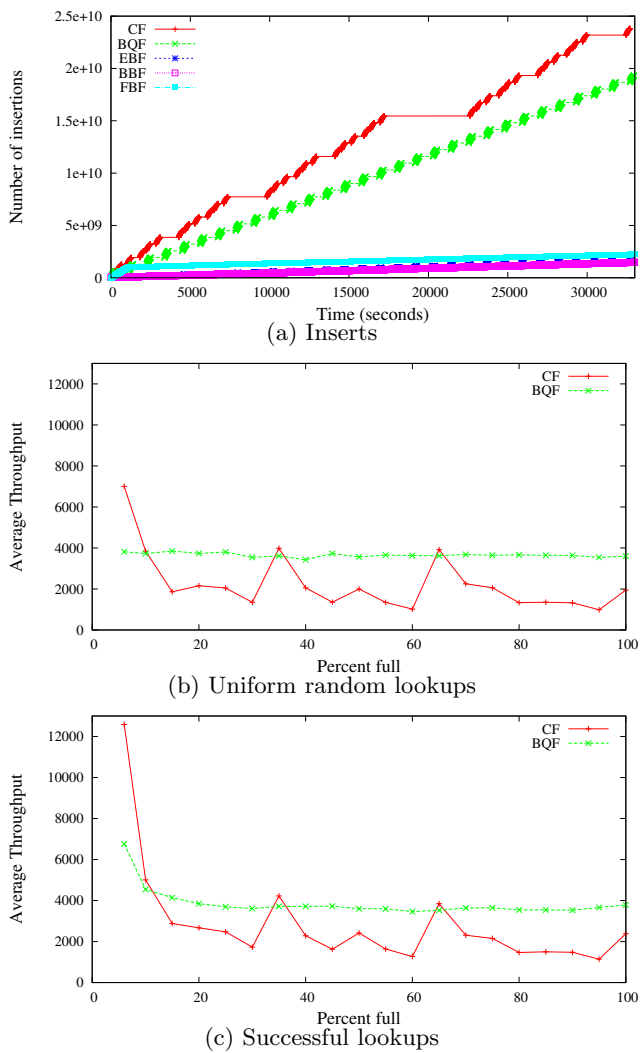

(a) Inserts

(b) Uniform random lookups

(c) Successful lookups

Figure 8: Large disk experiment. In 8(a), the cascade filter outperforms the buffered quotient filter. In 8(b) and 8(c) the cascade filter lookup performance depends on the number of levels it has: at 35 percent and 65 percent, it has only one level, and performs one random read, like buffered quotient filter.

The BQF outperforms the CF in the small experiment, but the CF outperforms the BQF in the large experiment, which is consistent with our prediction. Recall that an insert into the BQF requires $O(n/M/B)$ writes, and an insert into the CF requires $(O(\log(n/M)/B)$ writes. In the small experiment, $n/M \approx 4$, but in the large experiment, $n/M \approx 24$. Hence, the difference between $n/M$ and $\log(n/M)$ becomes significant and the CF begins to outperform the BQF. As the size of the database grows, the gap should get larger.

The insertion performance graphs also display the effects of each data structure's buffering strategy. For example, the stalls in the BQF performance correspond to flushing of the full in-RAM QF to the on-disk QF. The stalls become longer as the on-disk QF becomes fuller, making insertions into it more CPU-intensive. The stalls in the CF performance correspond to the merges of QFs. The largest stall is in the middle, where all but the in-RAM QFs are being merged into the largest QF in the CF. There are deamortization techniques, which we did not implement, that can remove such long stalls [1]. The EBF stalls during flushes, too, but each flush takes the same amount of time since BF insertion performance is independent of occupancy. The FBF insertion throughput starts high, during the FBF's in-RAM phase, but drops sharply once data begins spilling to disk. Although it appears to outperform the BBF and EBF in Figure 7(a), Table 1 shows that its overall performance is about 5x less than the BBF and EBF.

The EBF, BBF, and FBF were not able to complete the large experiment, so we cannot compare their overall performance, but we can report their performance on the insertions they completed. The FBF had a cumulative throughput of 67,000 insertions/second during the 35,000 second experiment. The BBF performed 44,600 inserts per second, and the EBF completed 53,000 insertions per second. The CF had a cumulative throughput of 728,000 insertions per second.

The lookup performance graphs support three conclusions: (1) the BQF and CF outperform the BF variants, (2) The BQF performs one random read per lookup, and (3) the CF performs between 1 and $\log(n/M)$ random reads per lookup. For uniform random lookups, the BQF performance is roughly 1.9 times higher than either the best BF variant or the CF. The CF uniform random lookup performance is comparable to the EBF and BBF performance, and almost 50% higher than the FBF uniform lookup rate. For successful lookups, the BQF performs 1.6 times better than the CF, 2.5 better than the FBF and 10 to 12 times better than the BBF and the EBF. The FBF maintains the most favorable successful lookup performance among the BF variants.

The EBF needs to perform $k = 12$ random reads for each successful lookup, which matches with our results. The BBF is slightly more efficient, due to hash localization and OS prefetching (the lookup indices are sorted.)

The CF always outperforms the BF variants, except under one circumstance. The FBF outperforms the CF when the CF has flushed to disk but the FBF is still operating in RAM. Since the FBF in-RAM phase uses a BF, which is slightly more space efficient than a QF, it can buffer more data before its first flush to disk. Hence the FBF outperforms the CF betwen 20% and 30% occupancy. Once the FBF flushes to disk, though, it becomes much slower than the CF. Also note that when both the CF and the FBF are operating in RAM, the CF is over twice as fast. Similarly, the BQF outperforms the BF variants once the structures have inserted 30% of the data.

The BQF and CF lookup performance curves match our theoretical analysis. The BQF performance is always around 4,000 lookups/second, consistent with the conclusion that each BQF lookup requires one random read and the empirical measurement that our SSD can perform about 4,000 random reads/second. The CF performance also matches theoretical predictions. For example, since the total data set size in the large experiment is 24 times larger than RAM, the CF should have between 1 and $4 \approx \log(24)$ active levels, and hence its lookup throughput should be between 1 and 4 times slower than the disk's random read throughput. Figures 8(b) and 8(c) match this expectation. The slowest points are at about 1,000 lookups/second, the fastest at 4,000 lookups/second.

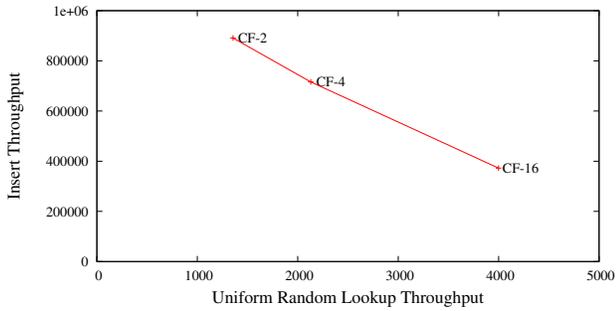

Figure 9: The Cascade Filter Insert/Lookup Tradeoff: Varying fanouts. Higher fanouts foster better lookup performance; lower fanouts optimize the insertion performance.

The lookup figures also reveal several other caching and buffering effects. Lookup throughputs for the CF and BQF exhibit a sawtooth pattern: the upside of the curve is due to populating the in-RAM QF, and thus satisfying a larger fraction of lookups in RAM. Throughput peaks right before the in-RAM QF is flushed – at 20%, 40% 60% and 80% in the small experiment. This effect is also more pronounced for successful lookups, since a successful lookup is more likely to stop in RAM. This effect becomes less significant as more data is inserted, since the data in the buffer becomes a smaller fraction of the inserted elements.

The BBF and the EBF uniform random lookup performance mildly decays as the data structures become fuller. This is due to the on-disk BF having more bits set to one as the occupancy of the filter grows. When the data structure is 100% full, the EBF and BBF need to check 2 bits on average. For EBF, this means 2 random reads; for the BBF, it is slightly less than 2 because two bits from the same subfilter (the erase block) may fall into the same read page. This is confirmed by our results, where the BBF slightly outperforms the EBF in lookups, but both are just above half of the random read throughput of the SSD.

### 5.3 Cascade Filter: Insert/Lookup Tradeoff

To investigate the effect of the fanout in the CF, we inserted 12 billion items into CFs with the same basic configuration as before: a 2GB buffer and a false positive rate of 1/4096. After inserting all 12 billion elements, we performed lookups for 60 seconds. We repeated this experiment with CFs for fanouts of 2, 4, and 16. Figure 9 shows the tradeoff between insert and lookup performance in these three experiments.

As expected, a higher fanout improves lookup performance, and a lower fanout improves insert performance. High fanouts reduce the number of levels in the CF, so lookups have fewer levels to check. The drawback of a high fanout is that each level will be written to disk several times, wasting disk bandwidth. According to Figure 9, even a fanout of 16 exceeds the insert performance of all the BF based data structures in our evaluations.

### 5.4 Evaluation Summary

QF-based data structures outperformed BF-based data structures in our evaluation. The QF outperforms the BF, although it uses more space in some configurations. The CF and BQF dramatically outperform all the BF variants. They can perform insertions an order of magnitude faster, and offer comparable or superior lookup performance.

The CF was the most scalable data structure in our experiments. As filter-to-RAM ratio grows, the CF outperforms the BQF. With ratios larger than 24, we expect the CF and the BQF performance to further diverge. When the ratio between the filter and the RAM buffer grows too large, then the flushes that BQF performs become distributed across the large filter, losing some of the space locality.

## 6. CONCLUSIONS

We have presented three efficient data structures – the quotient filter, the buffered quotient filter, and the cascade filter– for approximate membership testing. These data structures offer the same basic functionality as the well-known Bloom filter, but achieve much better data locality. The quotient filter uses slightly more space than the Bloom filter, but supports faster insertions, faster lookups, and deletes. The buffered quotient filter and cascade filter are efficient on-disk data structures built on top of the quotient filter. In our benchmarks, these structures offered insert performance an order of magnitude greater than recently-proposed Bloom filter based structures, and comparable or better lookup performance. The cascade filter also asymptotically scales better than other data structures, and this is reflected in our experimental results. Furthermore, the cascade filter was CPU-bound in our insertion benchmarks.

Future work could gain even more speed through parallelism. In our experiments, the cascade filter inserts used only serial writes and about 7MB/s of disk bandwidth (out of over 100MB/s of available bandwidth), so there is plenty of bandwidth to speed up inserts. We plan to focus on background merging, similar to the cleaner in log-structured filesystems, which could exploit idle disk bandwidth to improve lookup performance. Also, lookups are still hampered by slow random read disk performance: we plan to explore strategies to short-circuit lookups on non-existent elements. Finally, in this work we used only one Flash disk: our past work has shown performance variations across different disks and storage configurations [16]; we plan to explore the performance for different Flash disks and non-flash ones, as well as multi-disk systems such as RAID.

## 7. ACKNOWLEDGMENTS

Thanks to Guy Blelloch for helpful discussions and especially for suggesting that we use quotienting.

1636